\documentclass[final,5p,times,twocolumn]{elsarticle}

\usepackage[T1]{fontenc}
\usepackage[utf8]{inputenc}

\usepackage{amsmath}
\usepackage{amsfonts}
\usepackage{amssymb}
\usepackage{amsxtra}
\usepackage{array}
\usepackage{color}
\usepackage{dcolumn}
\usepackage{graphicx}
\usepackage{hepunits}
\usepackage{xspace}
\usepackage{CJKutf8}

\definecolor{purple}{rgb}{0.5,0,0.5}
\definecolor{blue}{rgb}{0.0,0,0.9}
\definecolor{prdblue}{rgb}{0.133,0.118,0.498}
\usepackage[colorlinks=true, pdfstartview=FitV, linkcolor=prdblue, citecolor= prdblue, urlcolor=prdblue]{hyperref}

\usepackage[mathscr,scaled=1.15]{urwchancal}
\DeclareFontFamily{OT1}{pzc}{}
\DeclareFontShape{OT1}{pzc}{m}{it}%
{<-> s * [1.15] pzcmi7t}{}
\DeclareMathAlphabet{\mathpzc}{OT1}{pzc}{m}{it}

\biboptions{sort&compress}

\journal{Physics Letters B}

\makeatletter

\setbox0\hbox{$\xdef\scriptratio{\strip@pt\dimexpr
    \numexpr(\sf@size*65536)/\f@size sp}$}

\newcommand{\scriptveryshortarrow}[1][3pt]{{%
    \hbox{\rule[\scriptratio\dimexpr\fontdimen22\textfont2-.2pt\relax]
               {\scriptratio\dimexpr#1\relax}{\scriptratio\dimexpr.4pt\relax}}%
   \mkern-4mu\hbox{\let\f@size\sf@size\usefont{U}{lasy}{m}{n}\symbol{41}}}}

\makeatother

\begin{document}
\begin{CJK}{UTF8}{song}

\begin{frontmatter}

\title{$\,$\\[-7ex]\hspace*{\fill}{\normalsize{\sf\emph{Preprint no}. NJU-INP 080/23}}\\[1ex]
Pion distribution functions from low-order Mellin moments}

\author[NJTU]{Ya Lu
    $^{\href{https://orcid.org/0000-0002-0262-1287}{\textcolor[rgb]{0.00,1.00,0.00}{\sf ID}},}$}

\author[UHe,UPO]{Yin-Zhen~Xu
       $^{\href{https://orcid.org/0000-0003-1623-3004}{\textcolor[rgb]{0.00,1.00,0.00}{\sf ID}},}$}

\author[UHe]{Kh\'epani Raya%
    $^{\href{https://orcid.org/0000-0001-8225-5821}{\textcolor[rgb]{0.00,1.00,0.00}{\sf ID}},}$}

\author[NJU,INP]{Craig D. Roberts%
       $^{\href{https://orcid.org/0000-0002-2937-1361}{\textcolor[rgb]{0.00,1.00,0.00}{\sf ID}},}$}

\author[UHe]{Jos\'e Rodr\'{\i}guez-Quintero%
       $^{\href{https://orcid.org/0000-0002-1651-5717}{\textcolor[rgb]{0.00,1.00,0.00}{\sf ID}},}$}

%
\address[NJTU]{
Department of Physics, Nanjing Tech University, Nanjing, Jiangsu 211816, China}

\address[UHe]{
Dpto.\ Ciencias Integradas, Centro de Estudios Avanzados en Fis., Mat. y Comp., Fac.\ Ciencias Experimentales, Universidad de Huelva, Huelva 21071, Spain}
\address[UPO]{Dpto. Sistemas F\'isicos, Qu\'imicos y Naturales, Univ.\ Pablo de Olavide, E-41013 Sevilla, Spain}
\address[NJU]{
School of Physics, Nanjing University, Nanjing, Jiangsu 210093, China}
\address[INP]{
Institute for Nonperturbative Physics, Nanjing University, Nanjing, Jiangsu 210093, China\\[1ex]
%
\href{mailto:cdroberts@nju.edu.cn}{cdroberts@nju.edu.cn} (C. D. Roberts);
\href{mailto:jose.rodriguez@dfaie.uhu.es}{jose.rodriguez@dfaie.uhu.es} (J. Rodr{\'{\i}}guez-Quintero)
\\[1ex]
Date: 2023 Nov 15\\[-6ex]
}

\begin{abstract}
Exploiting an evolution scheme for parton distribution functions (DFs) that is all-orders exact, contemporary lattice-QCD (lQCD) results for low-order Mellin moments of the pion valence quark DF are shown to be mutually consistent.   The analysis introduces a means by which key odd moments can be obtained from the even moments in circumstances where only the latter are available.  Combining these elements, one arrives at parameter-free lQCD-based predictions for the pointwise behaviour of pion valence, glue, and sea DFs, with sound uncertainty estimates.  The behaviour of the pion DFs at large light-front momentum fraction, $x\gtrsim 0.85$, is found to be consistent with QCD expectations and continuum analyses of pion structure functions, \emph{i.e}., damping like $(1-x)^{\beta_{\rm parton}}$, with
$\beta_{\rm valence} \approx 2.4$,
$\beta_{\rm glue} \approx 3.6$,
$\beta_{\rm sea} \approx 4.6$.
It may be possible to test these predictions using data from forthcoming experiments.
\end{abstract}

\begin{keyword}
continuum and lattice Schwinger function methods \sep
emergence of mass \sep
evolution equations \sep
parton distribution functions \sep
pion structure \sep
quantum chromodynamics
\end{keyword}

\end{frontmatter}
\end{CJK}


\section{Introduction}
%
As increasing investment is directed toward uncovering the origin of a nuclear size mass scale in Nature, \emph{i.e}, understanding the emergence of hadron mass \cite{Roberts:2021nhw, Binosi:2022djx, deTeramond:2022zcm, Salme:2022eoy, Ding:2022ows, Ferreira:2023fva, Carman:2023zke, Krein:2023azg}, there is a growing appreciation of the role that can be played by experimental studies of pion (and kaon) structure \cite{Adams:2018pwt, Aguilar:2019teb, Chen:2020ijn, Arrington:2021biu, Anderle:2021wcy, Quintans:2022utc, Wang:2023thl, Chavez:2023hbl, Accardi:2023chb}.
This emphasises the need for robust theoretical predictions of, \emph{inter alia}, pion distribution amplitudes (DAs) and functions (DFs).  Concerning DAs, modern predictions -- see, \emph{e.g}., Ref.\,\cite[Secs.\,3, 8\,D]{Roberts:2021nhw}, might be tested using the Drell-Yan process \cite{Brandenburg:1994wf, Xing:2023wuk}; and regarding DFs, analyses of existing data have been revisited \cite{Novikov:2020snp, Han:2020vjp, Barry:2021osv, Cui:2020tdf, Cui:2021mom, Chang:2023tob, Courtoy:2023bme} and many continuum and lattice studies have recently been completed -- see, \emph{e.g}., Refs.\,\cite{Roberts:2021nhw, 
Chang:2021utv, Raya:2021zrz, Lu:2022cjx, Lan:2021wok, dePaula:2022pcb, Cheng:2023kmt, Pasquini:2023aaf, Alexandrou:2021mmi, Joo:2019bzr, Sufian:2019bol, Gao:2022iex}.

It is worth recalling that a parton DF is a probability density distribution: ${\mathpzc p}^h(x;\zeta)\,dx$ is the number of partons within a hadron $h$ that carry a light-front fraction $x$ of the hadron's momentum when this is resolved at scale $\zeta$.  Each DF is an essentially nonperturbative quantity, relating directly to the wave function of the hadron \cite{Brodsky:1989pv, Mezrag:2023nkp}; so, charting DF $x$-dependence is one of the keys to understanding hadron structure.  In connection with the pion, almost all theory studies have focused on valence quark DFs, because they are the most straightforward.  Notwithstanding that, some analyses have recently tackled sea and glue DFs -- see, \emph{e.g}., Refs.\,\cite{Chang:2021utv, Fan:2021bcr, Bourrely:2022mjf, Pasquini:2023aaf}.

Models and continuum Schwinger function methods (CSMs) typically yield the full $x$-dependence of the DF \cite{Roberts:2021nhw, Holt:2010vj}.  Considering lattice-regularised QCD (lQCD), gaining access to the $x$-dependence once seemed an insurmountable problem; yet, today, methods have been proposed and are being developed for lQCD that also provide access to the functional dependence \cite{Ji:2013dva, Radyushkin:2017cyf, Ma:2017pxb}.  Issues remain, however, in some cases concerning the domain of support, which should be restricted to $x\in[0,1]$, and, in others, a need to solve or skirt an ``inverse problem''.  Consequently, a continuing focus of lQCD is the calculation of DF Mellin moments:
\begin{equation}
{\mathpzc M}_{\mathpzc p}^n(\zeta) =\langle x^n \rangle_{\mathpzc p}^\zeta = \int_0^1 dx\,x^n  {\mathpzc p}(x;\zeta).
\end{equation}

There are also challenges with calculating such Mellin moments \cite{Holt:2010vj}.  For instance, the ability to compute higher moments using lQCD is limited by statistical precision and, at a more basic level, by the breaking of $O(4)$ symmetry introduced by lattice discretisation.  In the calculation of higher-order moments when working with local operators, this introduces lattice-spacing power-divergences in the mixing with lower-dimensional operators, which restricts access to only those moments $\langle x^n \rangle$, $n\leq 3$ \cite{Alexandrou:2021mmi}.  Computation of higher-order moments is possible using nonlocal operators, in hybrid lQCD studies that exploit features of the frameworks introduced to make DF $x$-dependence available \cite{Joo:2019bzr, Sufian:2019bol}.  Notably, however, in some implementations of the Ioffe-time approach \cite{Radyushkin:2017cyf}, only even moments of the valence quark DF are accessible \cite{Gao:2022iex}.  This is an issue because the $n=1$ moment contains much important information, \emph{e.g}., it gives the momentum fraction carried by the valence quarks and can serve as the key to an evolution kernel between the subject DF at any two resolving scales \cite{Cui:2022bxn, Yin:2023dbw}.

Given a collection of lQCD calculations of some sets of low-order Mellin moments, a number of questions arise.
For instance:
having been obtained from distinct lattice setups, at different resolving scales, and using dissimilar algorithms, are they mutually consistent;
and supposing they are, is it possible to obtain a robust reconstruction of the DF, with reliable uncertainties,  from the available lQCD-determined Mellin moments?
Herein, we exemplify an approach to these questions and their answers using the pion valence quark DF moments reported in Refs.\,\cite{Alexandrou:2021mmi, Joo:2019bzr, Sufian:2019bol, Gao:2022iex}, which are listed in Table~\ref{latticemoments}.
Sketching briefly, these computations, respectively:
employ local operators to obtain low-order moments, with a practitioner-chosen fit used to infer higher moments;
reconstruct the DF from a lattice cross-section using a practitioner chosen fitting function, with low-order moments derived therefrom;
extract low-order moments from a pseudo-DF;
and employ a pseudo-DF scheme that only provides direct access to low-order even moments.
%

\begin{table}[t!]
\caption{
\label{latticemoments}
Lattice QCD results for Mellin moments of the pion valence-quark DF at
$\zeta=\zeta_2=2\,$GeV \cite{Joo:2019bzr, Gao:2022iex}
and
$\zeta_5=5.2\,$GeV \cite{Alexandrou:2021mmi, Sufian:2019bol}.
%
As discussed in connection with Eq.\,\eqref{EqGMin}, the column labelled ``G Eq.\,\eqref{EqGMin}'' provides the $\chi^2$ odd-moment completion of the Ref.\,\cite{Gao:2022iex} even moments.
}
\begin{tabular*}
{\hsize}
{
l@{\extracolsep{0ptplus1fil}}
|l@{\extracolsep{0ptplus1fil}}
l@{\extracolsep{0ptplus1fil}}
l@{\extracolsep{0ptplus1fil}}
l@{\extracolsep{0ptplus1fil}}
l@{\extracolsep{0ptplus1fil}}
l@{\extracolsep{0ptplus1fil}}}\hline\hline
$n\ $ & \cite[J]{Joo:2019bzr} & \cite[G]{Gao:2022iex} & G Eq.\,\eqref{EqGMin} &
\cite[A]{Alexandrou:2021mmi} & \cite[S]{Sufian:2019bol} \\\hline
$1\ $ & $0.254(03)\ $ & & $0.271\ $ & $0.23(1)\ $ & $0.18(3)\ $\\
$2\ $ & $0.094(12)\ $ & $0.1104(73)\ $ & &  $0.087(05)\ $ & $0.064(10)\ $\\
$3\ $ & $0.057(04)\ $ & & $0.054(8)\ $ & $0.041(04)\ $ & $0.030(05)\ $\\
$4\ $ & & $0.0388(46)\ $ &  & $0.023(05)\ $ & \\
$5\ $ & &  & $0.037(24)\ $ &  $0.014(04)\ $ & \\
$6\ $ & & $0.0118(48)\ $ & &   $0.009(03)\ $ & \\\hline\hline
%
%
%
%
%
%
%
\end{tabular*}
\end{table}

\section{DF evolution}
\label{SecAOE}
In discussing DFs, QCD evolution is crucial \cite{Dokshitzer:1977sg, Gribov:1971zn, Lipatov:1974qm, Altarelli:1977zs}.  We employ the all-orders scheme described succinctly in Ref.\,\cite[Sec.\,III]{Cui:2022bxn} and detailed in Ref.\,\cite{Yin:2023dbw}.  Here, to assist in making the presentation self-contained, we reiterate some points that are pertinent to the analysis of pion DF moments.

There are two primary principles.

\noindent {\sf P1} -- \emph{There exists at least one effective charge, $\alpha_{1\ell}(k^2)$, such that, when used to integrate the one-loop DGLAP equations, an evolution scheme for parton DFs is defined that is all-orders exact}.  

Charges of this type are reviewed in Ref.\,\cite{Deur:2023dzc}.  They need not be unique, but a suitable process-independent charge is not excluded.  That explained and calculated in Ref.\,\cite{Cui:2019dwv} has proved valuable, \emph{e.g}., serving to deliver a unified set of predictions for pion, kaon, and proton (unpolarised and polarised) DFs, and pion fragmentation functions that agree with much available data \cite{Cui:2020tdf, Lu:2022cjx, Cheng:2023kmt, Xing:2023pms}.
In being defined via observables, each such $\alpha_{1\ell}(k^2)$ is \cite{Deur:2023dzc}: consistent with the renormalisation group; renormalisation scheme independent; everywhere analytic and finite; and supplies an infrared completion of any standard running coupling.
It is, perhaps, worth highlighting that {\sf P1} does not require proof.  It is true by definition, as explained in Ref.\,\cite{Grunberg:1982fw}. 
The merits of such a scheme are judged by its efficacy.

\noindent{\sf P2} -- \emph{The hadron scale, $\zeta_{\cal H}<m_p$, where $m_p$ is the proton mass, is that scale at which valence (quasiparticle) degrees-of-freedom carry all properties of the given hadron, including, but not limited to, the entirety of its light-front momentum. $\zeta_{\cal H}$ is the initial scale for DF evolution.}

It follows from {\sf P2} that the glue and sea momentum fractions vanish at $\zeta_{\cal H}$ in every hadron; hence, since DFs are nonnegative on $x\in[0,1]$, ${\mathpzc g}^h(x;\zeta_{\cal H})\equiv 0 \equiv {\mathpzc S}^h(x;\zeta_{\cal H})$. 
To date, it has been found that the same value of $\zeta_{\cal H}$ serves well for all hadrons.  Regarding the pion, {\sf P2} entails
\begin{equation}
\label{Equpisymmetry}
{\mathpzc u}^\pi(x;\zeta_{\cal H})  = {\mathpzc u}^\pi(1-x;\zeta_{\cal H}) \,,
\quad
2 {\mathpzc M}_{{\mathpzc u}_\pi}^1(\zeta_{\cal H}) = 1\,.
\end{equation}
Kaon- and nucleon--like systems are discussed, respectively, in Refs.\,\cite{Raya:2021zrz, Lu:2022cjx}.

We now list three key corollaries of {\sf P1}, {\sf P2} for pion-like bound states \cite{Cui:2022bxn, Yin:2023dbw}.

\noindent{\sf C1} --  Since the hadron scale DF of a ground-state pseudoscalar meson is necessarily unimodal \cite[Sec.\,3]{Roberts:2021nhw}, then each moment or a realistic DF is bounded from above and below:
\begin{equation}
\label{momentbounds}
\frac{1}{2^n} \leq
{\mathpzc M}_{{\mathpzc u}_\pi}^n(\zeta_{\cal H})
%
\leq \frac{1}{1+n}\,,
\end{equation}
where the lower bound is provided by the moments of a bound state built from two infinitely heavy valence degrees-of-freedom and the upper expresses the moments of a pointlike system \cite{Cui:2022bxn}.

\noindent{\sf C2} --  Each moment of a DF at scale $\zeta$ is completely determined by the value of this moment at the hadron scale and the first moment at $\zeta$, \emph{viz}.\ for pion-like systems,
\begin{equation}
\label{EqxnzzH}
{\mathpzc M}_{{\mathpzc u}_\pi}^n(\zeta)
=
{\mathpzc M}_{{\mathpzc u}_\pi}^n(\zeta_{\cal H})
\left[2 {\mathpzc M}_{{\mathpzc u}_\pi}^1(\zeta)\right]^{\gamma_0^n/\gamma_0^1},
\end{equation}
where $\gamma_0^0=0$ and, for $n_f=4$ quark flavours, $\gamma_0^{1,2}=32/9, 50/9$.  The higher-$n$ results are listed elsewhere \cite[Eq.\,(6a)]{Yin:2023dbw}.  Hereafter, we write $\gamma_0^{n/1}=\gamma_0^n/\gamma_0^1$.

Using Eq.\,\eqref{EqxnzzH}, one obtains a form of Eq.\,\eqref{momentbounds} valid at any scale:
\begin{equation}
\label{momentbounds2}
\frac{1}{2^n}
\leq
{\mathpzc M}_{{\mathpzc u}_\pi}^n(\zeta) /
[ 2 {\mathpzc M}_{{\mathpzc u}_\pi}^1(\zeta)]^{\gamma_0^{n/1}}
\leq \frac{1}{1+n}\,.
\end{equation}

\noindent{\sf C3} -- Using Eqs.\,\eqref{Equpisymmetry}, \eqref{EqxnzzH}, one readily finds that each odd-order Mellin moment is completely determined by the set of lower-order even moments; hence,
\begin{align}
&
{\mathpzc M}_{{\mathpzc u}_\pi}^{2n+1}(\zeta)
= \frac{[2{\mathpzc M}_{{\mathpzc u}_\pi}^{1}(\zeta)]^{\gamma_0^{(2n+1)/1}}}
{2(n+1)}
\nonumber \\
& \times
\sum_{j=0,1,\ldots}^{2n}(-)^j
\left(\begin{array}{c} 2(n+1) \\ j \end{array}\right)
\frac{{\mathpzc M}_{{\mathpzc u}_\pi}^{j}(\zeta)}{
[2 {\mathpzc M}_{{\mathpzc u}_\pi}^{1}(\zeta)]^{\gamma_0^{j/1}}}.
\label{EqSymmB}
\end{align}
Any DF whose Mellin moments satisfy the Eq.\,\eqref{EqSymmB} recursion relation is linked by evolution to a symmetric distribution at $\zeta_{\cal H}$.  This is already known to be true \cite{Cui:2022bxn} for the lQCD studies reported in Refs.\,\cite{Alexandrou:2021mmi, Joo:2019bzr, Sufian:2019bol} and will herein be established for that in Ref.\,\cite{Gao:2022iex}.


\section{Odd moments from even}
\label{SecCompat}
It was shown elsewhere \cite{Cui:2022bxn} that the moments reported in Refs.\,\cite{Alexandrou:2021mmi, Joo:2019bzr, Sufian:2019bol} are mutually consistent and comply with all constraints described in Sec.\,\ref{SecAOE}.  It is now natural to enquire after the moments in Ref.\,\cite[G]{Gao:2022iex}.  This is a novel case because the method used therein only delivers even moments, with good signals for $n=2, 4, 6$ -- see Table~\ref{latticemoments}.

The question as to compatibility of the Ref.\,\cite[G]{Gao:2022iex} moments with the earlier studies can be addressed by considering the following $\chi^2$ measure:
\begin{subequations}
\label{chi2G}
\begin{align}
\chi^2_{\rm G}& =
\sum_{j={\rm A, J, S}}
\sum_{m=2}^{n_j} {\mathpzc a}_j^m
\frac{[{\mathpzc R}_{\,j}^m(\zeta_j)-{\mathpzc R}_{\,{\rm G}}^m(\zeta_{\rm G})]^2
}
{([\sigma_j^m]^2 + [\sigma_{\rm G}^m]^2) } \,,\\
{\mathpzc R}_{\,j}^m(\zeta_j) & = {\mathpzc M}_j^m(\zeta_j)/
[2{\mathpzc M}_j^1(\zeta_j)]^{\gamma_0^{m/1}} \,,
\end{align}
\end{subequations}
where we have dropped the ${\mathpzc u}_\pi$ subscript, because it is understood that only the pion is being discussed, replacing it with a label that indicates the source of the moment employed;
$n_j$ is the number of moments in lQCD study ``$j$'', with $\zeta_j$ being that study's resolving scale;
and ${\mathpzc a}_j^m = 1$ if moment $m$ is reported in study $j$ and zero otherwise, with
${\mathpzc M}_j^m$, $\sigma_j^m$ being the related moment and its uncertainty.
Since the ``G'' moments do not include $m=1$, this measure is presently ill-defined.

One way of proceeding is to introduce ${\mathpzc M}_{\rm G}^1(\zeta_{\rm G})$ as a parameter and find that value for this moment which minimises $\chi^2_{\rm G}$.  With that value in hand, the $n=3, 5$ moments can be obtained using Eq.\,\eqref{EqSymmB}.  Continuing in this way, one has a five-term minimisation with one fitting parameter, \emph{i.e}., four degrees of freedom; and the value
\begin{equation}
\label{EqGMin}
{\mathpzc M}_{\rm G}^1(\zeta_{\rm G}) = 0.271
\end{equation}
provides the minimum, with $\chi^2_{\rm G}/$degree-of-freedom$=0.95/4=0.24$.  This moment and the $n=3, 5$ moments obtained by recursion, Eq.\,\eqref{EqSymmB}, are also listed in Table~\ref{latticemoments}.  Standard error propagation methods entail that the uncertainty grows with $n$.  It can only be reduced by increasing the precision of the lower-order even moments.

It is here worth stressing that our odd-moment completion of the even moments reported in Ref.\,\cite{Gao:2022iex} makes neither assumptions about the form of the {\sf P1} effective charge nor the value of the {\sf P2} hadron scale.  It is therefore significant that using the PI charge elucidated in Ref.\,\cite{Cui:2019dwv}, denoted $\hat\alpha(k^2)$ and recorded explicitly elsewhere \cite[Eq.\,(13)]{Cui:2020tdf}, along with the first moment from Ref.\,\cite{Alexandrou:2021mmi}, one finds
\begin{subequations}
\label{EqConsistent}
\begin{align}
{\mathpzc M}_{\rm A}^1(\zeta_{\rm G}) & =
{\mathpzc M}_{\rm A}^1(\zeta_{\rm A})
\exp\left[-\frac{\gamma_0^1}{2\pi}\int_{\zeta_A}^{\zeta_G} \frac{dz}{z} \hat\alpha(z^2)\right] \label{Evolve}
\\
& = 0.269 (9)\,,
\end{align}
\end{subequations}
a value which agrees with the minimisation -- Eq.\,\eqref{EqGMin}.  It is further notable that repeating this procedure for all the moments reported in Ref.\,\cite{Gao:2022iex}, one arrives at the following comparisons:
\begin{equation}
\begin{array}{c|ccc}
n & 2 & 4 & 6 \\\hline
{\mathpzc M}_{\rm G}^n(\zeta_{\rm G}) & 0.1104(73) & 0.0388(46)  & 0.0118(48) \\
{\mathpzc M}_{\rm A}^n(\zeta_{\rm G}) & 0.102(11)\phantom{1}  & 0.027(09)\phantom{1}  & 0.011(05)\phantom{1}
\end{array}\,,
\end{equation}
which provide additional confirmation of both mutual consistency between lQCD results and validity of the analysis scheme described in Sec.\,\ref{SecAOE}.

\begin{figure}[t]
\centerline{\includegraphics[width=0.465\textwidth]{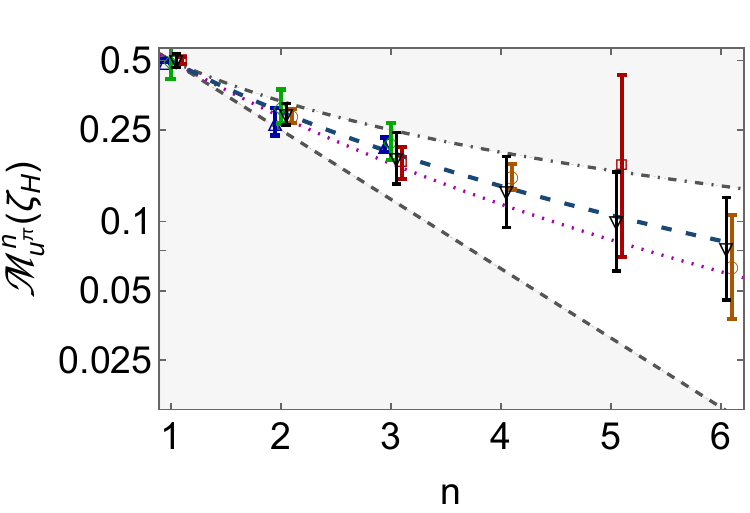}}
\caption{\label{PlotMoments}
Moments from Table~\ref{latticemoments}, referred to $\zeta_{\cal H}$ via Eq.\,\eqref{EqxnzzH}:
black down-triangles \cite[A]{Alexandrou:2021mmi};
blue up-triangles \cite[J]{Joo:2019bzr};
green diamonds \cite[S]{Sufian:2019bol};
orange circles  \cite[G]{Gao:2022iex} -- even moments; and
red squares  \cite[G]{Gao:2022iex} -- odd moments, obtained as described around Eq.\,\eqref{EqGMin}.
Results consistent with the bounds in Eq.\,\eqref{momentbounds2} fall within the open band.  The excluded regions are shaded lightly in grey.
%
%
Long-dashed dark-blue curve: moments of CSM DF \cite{Cui:2020tdf}, Eq.\,\eqref{CSMDF}.
Dotted magenta curve: moments of the scale-free distribution: ${\mathpzc q}^{\rm sf}(x)=30 x^2(1-x)^2$.
}
\end{figure}

In Fig.\,\ref{PlotMoments}, we depict all moments in Table~\ref{latticemoments}, evolved to the hadron scale using Eq.\,\eqref{EqxnzzH}.  Evidently, all considered lQCD studies deliver moments that are mutually consistent and satisfy the physical bounds, Eq.\,\eqref{momentbounds}.

Whilst immaterial for the comparisons discussed, it is nevertheless interesting to identify a typical value of the hadron scale which may be associated with each lQCD simulation.  This can be achieved by using the PI charge \cite[Eq.\,(13)]{Cui:2020tdf} to find that value of $\zeta_{\cal H}$ for which Eq.\,\eqref{EqxnzzH} yields ${\mathpzc M}_j^1(\zeta_{\cal H}) =1/2$ from the first moment associated with simulations $j=\,$A, J, S, G at their respective resolving scales, $\zeta_j$.  Ignoring uncertainties, this procedure yields
\begin{equation}
\label{Eqzetavalues}
\begin{array}{c|cccc}
{\rm lQCD~study}   & \mbox{\cite[A]{Alexandrou:2021mmi}} & \mbox{\cite[J]{Joo:2019bzr}} & \mbox{\cite[S]{Sufian:2019bol}} & \mbox{\cite[G]{Gao:2022iex}} \\\hline
\zeta_{\cal H}/{\rm GeV} & 0.379 & 0.351 & 0.287 & 0.381 \\
\end{array}\,.
\end{equation}
These values are well clustered, with a mean $0.350(44)\,$GeV that is consistent with the CSM prediction \cite[Eq.\,(15)]{Cui:2020tdf}: $\zeta_{\cal H} =0.331(2)\,$GeV.

The moments of the CSM prediction:
\begin{equation}
{\mathpzc u}^\pi(x;\zeta_{\cal H}) =
{\mathpzc n}_\pi \ln[ 1+x^2(1-x)^2/\rho^2]\,,
\label{CSMDF}
\end{equation}
with $\rho = 0.0660$ and ${\mathpzc n}_\pi$ a constant that ensures unit normalisation, are also shown in Fig.\,\ref{PlotMoments} -- long-dashed blue curve: within uncertainties, all lQCD results are consonant with this curve.
Notably, the function in Eq.\,\eqref{CSMDF} is flexible enough to simultaneously express both the dilation that the phenomenon of emergent hadron mass is known to generate in the valence quark DF \cite{Lu:2022cjx} and endpoint ($x\simeq 0,1$) behaviour matching QCD expectations -- see, \emph{e.g}., Ref.\,\cite[Sec.\,V]{Cui:2022bxn} and Sec.\,\ref{SecLargeX} below.

\begin{figure}[t]
\centerline{\includegraphics[width=0.465\textwidth]{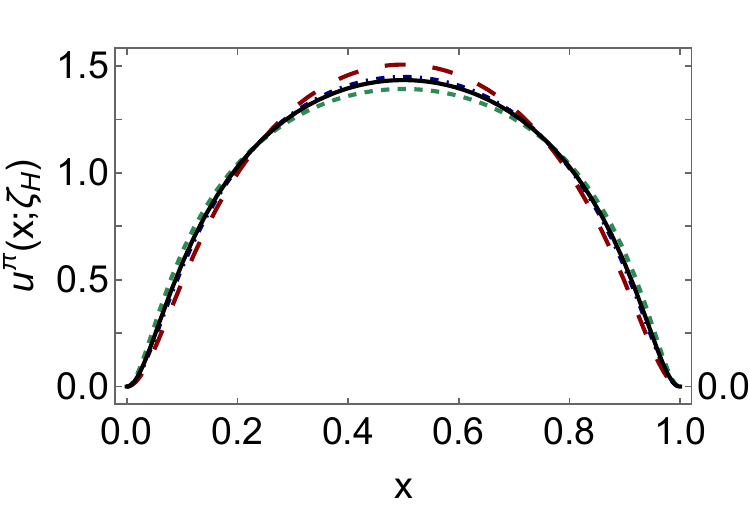}}
\caption{\label{FigDFszH}
Referred to the one-parameter function in Eq.\,\eqref{CSMDF}, pion DF obtained by requiring a best $\chi^2$ fit to the $n\geq 2$ moments in Refs.\,\cite{Alexandrou:2021mmi, Joo:2019bzr, Sufian:2019bol} -- dashed dark-green curve;
$n=2,\ldots,6$ moments associated with Ref.\,\cite{Gao:2022iex} -- long-dashed dark-red curve;
all $n\geq 2$ moments in Table~\ref{latticemoments} -- solid black curve.
The CSM prediction is drawn as the dot-dashed dark-blue curve.
}
\end{figure}

At this point, one may ask for a $\chi^2$ best-fit to all lQCD moments expressed through a value of $\rho$ in Eq.\,\eqref{CSMDF}.
Using only the $n\geq 2$ moments from Refs.\,\cite{Alexandrou:2021mmi, Joo:2019bzr, Sufian:2019bol}, the usual $\chi^2$ function is minimised by \cite{Cui:2022bxn} $\rho=0.048$, with $\chi^2/$degree-of-freedom$\,=0.27$.
(\emph{N.B}.\  Eq.\,\eqref{CSMDF} is a symmetric distribution and ${\mathpzc M}^1(\zeta_{\cal H})=1/2$ for all studies; hence,  only $n\geq 2$ moments are relevant in the minimisation.)
Focusing instead solely on the $n=2,\ldots,6$ moments associated with Ref.\,\cite{Gao:2022iex}, one finds $\rho=0.088$ with $\chi^2/$degree-of-freedom$\,=2.2/4=0.55$.
Combining all $n\geq 2$ moments listed in Table~\ref{latticemoments},
$\rho=0.061$ with $\chi^2/$degree-of-freedom$\,4.8/13=0.37$.
The hadron scale DFs obtained with these values of $\rho$ are drawn in Fig.\,\ref{FigDFszH}:
the mean ${\mathpzc L}_1$ difference between all curves drawn is $4.1(2.3)$\%; namely, they are practically indistinguishable.  The ${\mathpzc L}_1$ difference between the combined lQCD result and the CSM prediction is $1.0$\%.
(Since all valence DFs bound unit area, this measure is simply the integral of the absolute value of the difference between the curves.)

\section{Pion DFs from lattice-QCD moments}
\label{SecDFs}
We have seen that each of the simulations represented in Table~\ref{latticemoments} is in accord with the valence quark DF features explained in Sec.\,\ref{SecAOE}, \emph{viz}.\ {\sf P1}, {\sf P2}, and their corollaries; and that these lQCD studies are all mutually consistent.  Hence, one may combine the moments in Table~\ref{latticemoments} to obtain an optimal description of the entire collection.

This can be accomplished by first considering the hadron-scale DF in Eq.\,\eqref{CSMDF}.
Then, denoting the moments of this function by ${\mathpzc M}_\pi^n(\rho)$, one minimises the following uncertainty-weighted $\chi^2$ measure:
\begin{equation}
\chi^2(\rho) = \!\!\!
\sum_{j={\rm A, J, S, G}}\,
\sum_{n=2}^6 a_{j}^n
\frac{ [{\mathpzc M}_\pi^n(\rho) - {\mathpzc R}_{\,j}^n(\zeta_j)]^2}
{[\sigma_{j}^n]^2}\,, \label{X2distribution}
\end{equation}
where $a_{j}^n=1$ in all cases with an entry in Table~\ref{latticemoments} and is otherwise zero; and
$M^{\rm s}_n(\zeta)$, $\sigma^{\rm s}_n$ are the related moment and uncertainty.
As reported above, this yields $\rho_0=0.061$, $\chi^2(\rho_0) =4.8/13 = 0.37$.
(The  uncertainties in Eq.\,\eqref{X2distribution} are subsequently rescaled such that $\chi_0^2:=\chi^2(\rho_0) = d-2$, where $d=13$.)

\begin{figure}[t]
\vspace*{2ex}

\leftline{\hspace*{0.5em}{\large{\textsf{A}}}}
\vspace*{-3ex}
\includegraphics[width=0.465\textwidth]{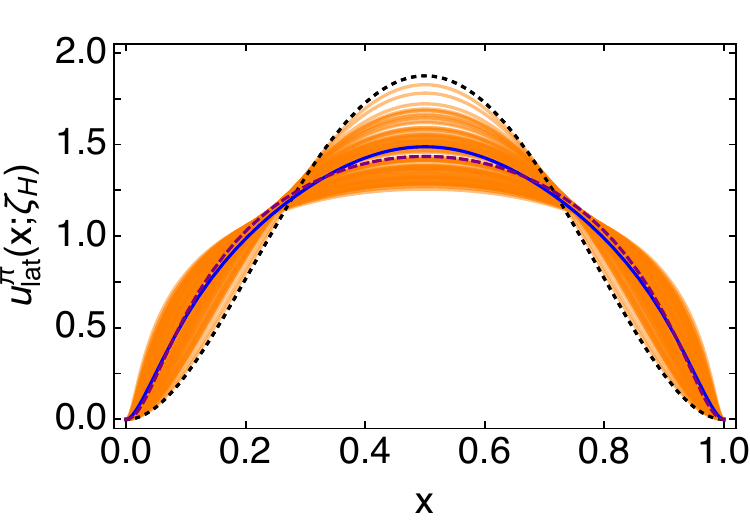}
\vspace*{-1ex}

\leftline{\hspace*{0.5em}{\large{\textsf{B}}}}
\vspace*{-3ex}
\includegraphics[width=0.465\textwidth]{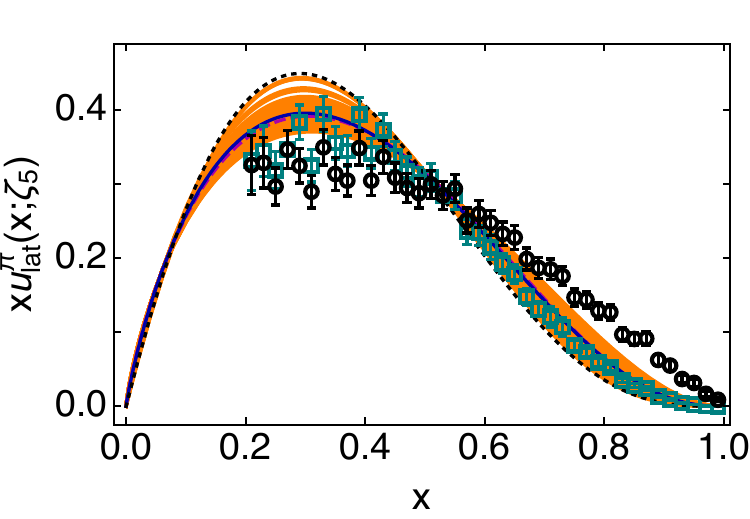}
\caption{\label{Fensemble}
\emph{Upper panel}\,--\,{\sf A}.  Randomly distributed ensemble of lQCD-based -- see Table~\ref{latticemoments} -- valence-quark DFs (orange curves) constructed using the procedure described in connection with Eq.\,\eqref{EqProb}.
\emph{Lower panel}\,--\,{\sf B}.  $\zeta_{\cal H}\to \zeta_{5}$ evolution of each curve in Panel {\sf A}. 
Black circles, data recorded in Ref.\,\cite[E615]{Conway:1989fs};
and teal boxes, reevaluation of that data as presented in Ref.\,\cite{Aicher:2010cb}.
Both panels.
Dashed purple curve: central $\rho=\rho_0$ result in Eq.\,\eqref{CSMDF}.
Solid blue curve: CSM prediction from Ref.\,\cite{Cui:2020tdf}.
Dotted black curve: scale-free distribution.
(All at scale appropriate to the panel.)
}
\end{figure}

Exploiting this result, we generate a set of curves that express the uncertainty in the lQCD moments as follows.
(\emph{a}) From a distribution centred on $\rho_0$, choose a new value of $\rho$.
(\emph{b})
Evaluate $\chi^2(\rho)$ in Eq.\,\eqref{X2distribution}.
The new value of $\rho$ is accepted with probability
\begin{equation}
\label{EqProb}
{\mathpzc P}  = \frac{P(\chi^2;d)}{P(\chi_0^2;d)} \,, \;
P(y;d) = \frac{(1/2)^{d/2}}{\Gamma(d/2)} y^{d/2-1} {\rm e}^{-y/2}\,.
\end{equation}
%
%
(\emph{c}) Repeat (\emph{a}) and (\emph{b}) until one has a $K \gtrsim 100$-member set of hadron-scale DFs.
(One can use more, but the impact is immaterial.)
This procedure yields the ensemble of DFs drawn in Fig.\,\ref{Fensemble}A.
Plainly, they properly bracket the central curve, which, itself, is a close match to the CSM prediction \cite{Cui:2020tdf}.

Capitalising on {\sf P1}, each curve in Fig.\,\ref{Fensemble}A can be evolved to $\zeta_5$ once
$2{\mathpzc M}_{{\mathpzc u}^\pi}^1(\zeta_5)$ is known -- see Eq.\,\eqref{EqxnzzH}.
Using an uncertainty weighted average of the results in Table~\ref{latticemoments}, \emph{viz}.\
$2{\mathpzc M}_{{\mathpzc u}^\pi}^1(\zeta_5) = 0.438(5)$, obtained after the $j=\,$J, G $\zeta_2$ values were evolved to $\zeta_5$ via analogues of Eq.\,\eqref{Evolve},
and no further information, one obtains the orange curves in Fig.\,\ref{Fensemble}B.
The central curve and associated $1\sigma$-band are reproduced by
\begin{equation}
\label{lQCDexponents}
{\mathpzc u}^\pi(x;\zeta_5)
= {\mathpzc n}_0^{\zeta_5} x^\alpha (1-x)^\beta (1+\gamma x^2)\,,
\end{equation}
$\alpha=-0.134(62)$, $\beta=2.55(36)$, $\gamma=1.62(77)$,
with ${\mathpzc n}_0^{\zeta_5}$ ensuring unit normalisation.
Evidently, the lQCD results are consistent with the reanalysis of E615 data described in Ref.\,\cite{Aicher:2010cb}, which follows the so-called Mellin-Fourier approach to resummation of next-to-leading-logarithms that was also employed with similar effect in Ref.\,\cite{Barry:2021osv}.  

\begin{figure}[t]
\centerline{\includegraphics[width=0.46\textwidth]{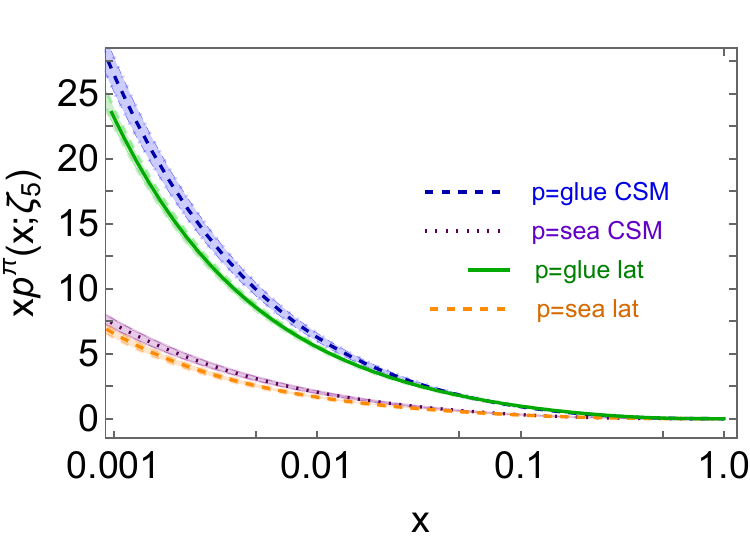}}
%
\caption{\label{PlotGlueSea}
Glue and sea DFs at $\zeta_5=5.2\,$GeV.  The band associated with each curve expresses consequences of the uncertainty in the valence momentum fraction:
$2 {\mathpzc M}_{{\mathpzc u}_\pi}^1(\zeta_5):= 0.438(5)$;
leading to
${\mathpzc M}_{{\mathpzc g}_\pi}^1(\zeta_5):= 0.436(1)$;
${\mathpzc M}_{{\mathpzc S}_\pi}^1(\zeta_5):= 0.125(1)$.
For comparison, CSM predictions from Refs.\,\cite{Cui:2020tdf, Chang:2021utv} are also drawn: in this case,
$2 {\mathpzc M}_{{\mathpzc u}_\pi}^1(\zeta_5):= 0.40(2)$;
${\mathpzc M}_{{\mathpzc g}_\pi}^1(\zeta_5):= 0.45(1)$;
${\mathpzc M}_{{\mathpzc S}_\pi}^1(\zeta_5):= 0.14(1)$.
}
\end{figure}

Exploiting {\sf P1}, then the results in Fig.\,\ref{Fensemble}A also enable prediction of pion glue and sea DFs \cite{Yin:2023dbw}.  Employing the central curve in Fig.\,\ref{Fensemble}A, generated with $\rho=\rho_0 = 0.061$ in Eq.\,\eqref{CSMDF}, one obtains the DFs in Fig.\,\ref{PlotGlueSea}.  On the entire kinematic domain, there is fair agreement between the lQCD-based results calculated herein and the CSM predictions \cite{Cui:2020tdf, Chang:2021utv}.  It is worth stressing that the CSM result for the glue DF \cite{Chang:2021utv} agrees with an independent lattice determination \cite{Fan:2021bcr}; hence, the lQCD-based result calculated herein is also in accord with that study.

\section{Pion DFs at large-$\mathbf x$}
\label{SecLargeX}
Analyses of the pion valence-quark DF, which incorporate the behaviour of the pion wave function prescribed by QCD, predict \cite{Cui:2021mom}:
\begin{equation}
\label{pionDFpQCD}
{\mathpzc u}^\pi(x;\zeta) \stackrel{x\simeq 1}{\sim} (1-x)^{\beta \,=\,2+\gamma(\zeta)}\,,
\end{equation}
where $\gamma(\zeta_{\cal H})=0$ and $\gamma(\zeta>\zeta_{\cal H}) \geq 0$ grows logarithmically with $\zeta$, expressing the physics of gluon radiation from the struck quark.  The powers on glue and sea DFs are, respectively, one and two units greater \cite{Brodsky:1994kg, Yuan:2003fs, Holt:2010vj, Cui:2021mom, Cui:2022bxn, Lu:2022cjx}.
Nevertheless, long after the first experiment relevant to ${\mathpzc u}^\pi(x\simeq 1)$ \cite{Corden:1980xf}, data-based conclusions relating to these predictions remain confused because, amongst the many methods used to fit existing data, \emph{e.g}., Refs.\,\cite{Aicher:2010cb, Novikov:2020snp, Han:2020vjp, Barry:2021osv, Cui:2020tdf, Cui:2021mom, Chang:2023tob, Courtoy:2023bme}, some produce a ${\mathpzc u}^\pi$ form that violates Eq.\,\eqref{pionDFpQCD} and its corollaries. The results in Fig.\,\ref{Fensemble} bear directly upon this issue.

As noted in connection with Eq.\,\eqref{lQCDexponents}, combined, the recent lQCD analyses collected in Table~\ref{latticemoments} produce a large-$x$ exponent $\beta(\zeta_5) = 2.55 (36)$.  However, this is the exponent on $x\simeq 1$, a domain whereupon data cannot readily be obtained.  For empirical purposes, it is more useful to report an effective exponent, \emph{i.e}., a slope parameter averaged over the domain $x\in [0.85,1.0]$ \cite{Holt:2010vj}.  Working with the results in Fig.\,\ref{Fensemble}B, one finds
\begin{equation}
  \beta_{\rm valence}^{\rm eff}(\zeta_5) \stackrel{x \in [0.85,1]}{=} 2.36(31)\,.
\end{equation}
Restricting the domain to $x\in [0.9,1.0]$, this value rises 2\%, becoming $2.41(31)$.
Analogous results for glue and sea are:
\begin{subequations}
\label{ExponentsGlueSea}
\begin{align}
  \beta_{\rm glue}^{\rm eff}(\zeta_5) & \stackrel{x \in [0.85,1]}{=} 3.56(23)\,,\\
  \beta_{\rm sea}^{\rm eff}(\zeta_5) & \stackrel{x \in [0.85,1]}{=} 4.62(26)\,.
\end{align}
\end{subequations}
Modern lQCD results are thus seen to be consistent with Eq.\,\eqref{pionDFpQCD} and its corollaries.

\section{Summary and outlook}
Contemporary simulations of lattice-regularised QCD (lQCD) produce Mellin moments associated with the pion valence quark distribution function (DF) that are consistent with an array of expectations based on the all-orders evolution scheme [Sec.\,\ref{SecAOE}].
Seen from this perspective, they are also mutually compatible [Sec.\,\ref{SecCompat}].
Consequently, they may be combined to deliver parameter-free lQCD-based predictions for the pointwise behaviour (light-front momentum fraction, $x$, dependence) of pion valence, glue, and sea DFs, with quantitatively reliable uncertainties [Sec.\,\ref{SecDFs}].
Consistent with modern continuum predictions at the resolving scale usually associated with E615 data \cite{Conway:1989fs}, the large-$x$ behaviour of the lQCD DFs can be represented via [Sec.\,\ref{SecLargeX}]
\begin{equation}
(1-x)^{\beta_{\mathpzc p}^{\rm eff}},
\end{equation}
with
$\beta_{\mathpzc u}^{\rm eff} \approx 2.4$,
$\beta_{\mathpzc g}^{\rm eff} \approx 3.6$,
$\beta_{\mathpzc S}^{\rm eff} \approx 4.6$.
These predictions can both serve as benchmarks for existing data fitting methods and, once those methods are shown to be reliable, be tested using data from forthcoming experiments.
Moreover, using crossing symmetry \cite{Drell:1969jm, Gribov:1972ri, Gribov:1972rt}, they can be used to develop lQCD-based predictions for pion fragmentation functions \cite{Xing:2023pms}.

%
\medskip
\noindent\textbf{Acknowledgments}.
We are grateful to Z.-F.~Cui and J.-H.~Zhang for valuable discussions.
Work supported by:
National Natural Science Foundation of China (grant nos.\,12135007, 12205149);
Spanish Ministry of Science and Innovation (MICINN) (grant no.\ PID2022-140440NB-C22);
Junta de Andaluc{\'{\i}}a (grant no.\ P18-FR-5057);
and
STRONG-2020 ``The strong interaction at the frontier of knowledge: fundamental research and applications” which received funding from the European Union's Horizon 2020 research and innovation programme (grant agreement no.\ 824093).

\medskip
\noindent\textbf{Data Availability Statement}. This manuscript has no associated data or the data will not be deposited. [Authors' comment: All information necessary to reproduce the results described herein is contained in the material presented above.]

\medskip
\noindent\textbf{Declaration of Competing Interest}.
The authors declare that they have no known competing financial interests or personal relationships that could have appeared to influence the work reported in this paper.


\end{document}